\documentclass[prd,amsmath,amssymb,twocolumn]{revtex4}
\usepackage{graphicx}

\def\al{\alpha}
\def\be{\beta}

\def\de{\delta}
\def\ep{\epsilon}

\def\ka{\kappa}
\def\la{\lambda}

\def\mn{{\mu\nu}}

\def\prt{\partial}
\def\cl{{\cal L}}

\def\fr#1#2{{{#1} \over {#2}}}
\def\Frac#1#2{{\textstyle{{#1}\over {#2}}}}
\def\half{{\textstyle{1\over 2}}}
\def\lsim{\mathrel{\rlap{\lower4pt\hbox{\hskip1pt$\sim$}}
    \raise1pt\hbox{$<$}}}
\def\gsim{\mathrel{\rlap{\lower4pt\hbox{\hskip1pt$\sim$}}
    \raise1pt\hbox{$>$}}}
\def\sqr#1#2{{\vcenter{\vbox{\hrule height.#2pt
         \hbox{\vrule width.#2pt height#1pt \kern#1pt
         \vrule width.#2pt}
         \hrule height.#2pt}}}}
\newcommand{\beq}{\begin{equation}}
\newcommand{\eeq}{\end{equation}}
\newcommand{\bea}{\begin{eqnarray}}
\newcommand{\eea}{\end{eqnarray}}
\newcommand{\rf}[1]{(\ref{#1})}

\def\mbf#1{\mbox{\boldmath$#1$}}
\def\Re{{\rm Re}}
\def\kde{\ka_{DE}}
\def\kdb{\ka_{DB}}
\def\khe{\ka_{HE}}
\def\khb{\ka_{HB}}
\def\kep{\tilde\ka_{e+}}
\def\kem{\tilde\ka_{e-}}
\def\kop{\tilde\ka_{o+}}
\def\kom{\tilde\ka_{o-}}
\def\ktr{\tilde\ka_{\rm tr}}
\def\ede{\ep_{DE}}
\def\edb{\ep_{DB}}
\def\ehe{\ep_{HE}}
\def\ehb{\ep_{HB}}
\def\Mep{{\cal\widetilde M}_{e+}}
\def\Mem{{\cal\widetilde M}_{e-}}
\def\Mop{{\cal\widetilde M}_{o+}}
\def\Mom{{\cal\widetilde M}_{o-}}
\def\Mtr{{\cal\widetilde M}_{\rm tr}}
\def\Mde{{\cal M}_{DE}}
\def\Mhb{{\cal M}_{HB}}
\def\Mdb{{\cal M}_{DB}}

\begin{document}

\title{Cavity tests of parity-odd Lorentz violations in electrodynamics}
\author{Matthew Mewes}
\affiliation{Physics Department, Marquette University,
  Milwaukee, WI 53201, U.S.A.}
\author{Alexander Petroff}
\affiliation{Department of Earth, Atmospheric, and Planetary Sciences,
  Massachusetts Institute of Technology,
  Cambridge, MA 02139, U.S.A.}
\date{\today} 


\begin{abstract}
Electromagnetic resonant cavities
form the basis for a number modern
tests of Lorentz invariance.
The geometry of most of these experiments
implies unsuppressed sensitivities to
parity-even Lorentz violations only.
Parity-odd violations typically
enter through suppressed boost effects,
causing a reduction in sensitivity
by roughly four orders of magnitude.
Here we discuss possible techniques for
achieving unsuppressed sensitivities
to parity-odd violations by using
asymmetric resonators.
\end{abstract}
\maketitle

\section{introduction}

In recent years,
renewed interest in 
precision tests of relativity 
has resulted in a number of 
modern version of the classic
Michelson-Morley
\cite{mm} and 
Kennedy-Thorndike
\cite{kt}
experiments
\cite{cavities}.
These tests are motivated
in part by the observation that
attempts to quantize 
gravity may lead to tiny
violations of Lorentz invariance
at attainable energies.
While originally conceived within
the context of spontaneous symmetry 
breaking in string theory
\cite{ks,kp},
a number of other possible origins
have been proposed
\cite{ncqed,qg,fn,bj}.
Remarkably,
naive estimates suggest that
these violations may be within 
reach of contemporary experiment
\cite{cpt,ck,kost}.

Lorentz invariance includes
covariance under
both rotations and boosts.
Traditionally,
Michelson-Morley-type experiments
test the rotational invariance,
while Kennedy-Thorndike experiments
focus on boost symmetry.
Modern versions are normally
sensitive to both types of violations,
but sensitivities to boost effects
are usually suppressed relative to
rotational violations due to the
small velocities involved in
most experiments.
The symmetry of most resonators
imply that only parity-even
violations of Lorentz invariance
are observable in Michelson-Morley tests
and are detectable
at unsuppressed levels.
Parity-breaking and isotropic
Lorentz violations enter at
first and second
order in small velocities,
causing reduced sensitivities.

While sensitivities to Lorentz violations
in photons continue to improve,
a substantial increase in
sensitivity to parity-odd violations
may be possible in experiments that
do not respect parity symmetry.
In this work,
we focus on resonator experiments,
exploring the reasons behind
the suppression and the potential
for parity-asymmetric resonators
to yield higher sensitivities
to parity-odd Lorentz violations.
Other suggestions for improving
sensitivity to parity-odd 
violations include
searches for mixing between
the electric-field vector and
the magnetic-field pseudovector in
electromagnetostatic experiments
\cite{kb}
and the use of interferometers
or traveling-wave resonators
\cite{tobar}.

General violations of Lorentz
invariance are described by
a field-theoretic framework
known as the Standard-Model Extension (SME)
\cite{ck,kost}.
The SME provides a systematic
theoretical basis for studies of
Lorentz invariance in many systems,
including those involving photons
\cite{cavities,km,tobar,kb,km-bire,cfj,photonth,sher},
baryons \cite{ccexpt,spaceexpt},
hadrons \cite{hadrons,hadronth},
electrons \cite{electrons,electronth},
muons \cite{muons},
neutrinos \cite{neutrinos},
Higgs bosons \cite{higgs},
and
gravitation \cite{kb-gr,kp2}.
Here we work within the 
renormalizable gauge-invariant
$CPT$-even photon sector of the 
minimal SME,
but many of the symmetry arguments
presented here may be extended
to more general violations.

The structure of this paper
is as follows.
Section \ref{framework}
gives a review of the
theory and notation
used in this paper.
The general theory behind
resonator experiments
is described in
Sec.\ \ref{resonators}.
The possibility of using
resonators with parity-breaking
geometries is discussed in
Sec.\ \ref{parity-breaking}.
Section \ref{example}
presents a numerical example
of a parity-asymmetric resonant
cavity.
Some concluding remarks are
given in Sec.\ \ref{discussion}.

\section{basic theory} \label{theory}

This section provides some
basic theory and definitions.
Lorentz violation in the
photon sector of the minimal 
SME is reviewed,
and the characterization of potential
sensitivities in general resonator
experiments is given.

\subsection{Framework}
\label{framework}

The violations of interest
are described
by a modified Maxwell lagrangian
\cite{ck},
\beq
\cl = -\frac14 F^\mn F_\mn
-\frac14 (k_F)^{\ka\la\mu\nu}
F_{\ka\la} F_{\mu\nu}\ ,
\label{lagr}
\eeq
where tensor coefficients 
$(k_F)^{\ka\la\mu\nu}$
characterize the extent
to which Lorentz symmetry is violated.
The tensor $(k_F)^{\ka\la\mu\nu}$
is real and obeys the symmetries
of the Riemann tensor.
In addition,
the double trace is usually
assumed to be zero since
it only contributes to
a scaling of the theory.
This leaves a total of
19 independent coefficients
for Lorentz violation.
These coefficients are taken
to be constant in the minimal SME, 
but may depend on spacetime location
in more general contexts,
including those involving
gravitation \cite{kost,kp2,kb-gr}.
Other forms of Lorentz violation
that could also be considered
include the $CPT$-odd $k_{AF}$
term of the minimal SME \cite{ck},
and nonrenormalizable terms
of the general SME \cite{kpo,kle}.

The equations of motion associated
with lagrangian \rf{lagr} provide
modified inhomogeneous Maxwell equations.
It turns out that these can
be cast into the familiar form
$\mbf\nabla\times\mbf H -\prt_0\mbf D = 0$,
$\mbf\nabla\cdot\mbf D = 0$, 
provided we define
\cite{km}
\begin{align}
  \mbf D =\ & (\ede+\kde)\cdot\mbf E 
  + (\edb+\kdb)\cdot\mbf B\ , \notag \\
  \mbf H =\ & (\ehe+\khe)\cdot\mbf E 
  + (\ehb+\khb)\cdot\mbf B\ .
  \label{const}
\end{align}
Here we allow for
the possibility of general linear
passive magnetoelectric media
with constituent matrices
$\ede$,
$\edb$,
$\ehe$, and
$\ehb$
\cite{kong}.
For harmonic fields,
these matrices are complex
and may depend on frequency.
Losslessness implies that 
$\ede$ and $\edb$ are hermitian
and $\ehe=-\edb^\dag$.
In many applications these reduce
to a simple isotropic permittivity
and permeability:
$\ep_{DE} = \ep$,
$\ep_{HB}=\mu^{-1}$, and 
$\ep_{DB} = \ep_{HE} = 0$.

In this language,
Lorentz violation in photons
is controlled by the real
$3\times 3$ matrices
$\kde$,
$\kdb$,
$\khe$, and
$\khb$,
which result from a 1-3 decomposition
of the $(k_F)^{\ka\la\mu\nu}$ tensor.
These matrices obey the same
lossless conditions
as their $\ep$ counterparts.
Note that $\kde$ and $\khb$
are parity conserving,
while $\kdb=-\khe^T$ mixes vectors
and pseudovectors,
introducing parity violations.
Also note that it is usually 
assumed that the $\ep$ matrices
are not significantly altered
by Lorentz violation in photons.

A subset of the coefficients for
Lorentz violation cause vacuum birefringence,
which can be tested with extreme precision 
by polarimetry of light from sources
at cosmological distances
\cite{cfj,km-bire,km}.
It is therefore useful to
decompose the $\ka$ matrices into
coefficients that cause birefringence
and those that do not:
\begin{align}
  \kde &= \kep+\kem+\ktr \ , \notag \\
  \khb &= \kep-\kem-\ktr \ , \\
  \kdb &= -\khe^T=\kop+\kom \ . \notag
\end{align}
Here
$\kep$,
$\kem$,
$\kop$, and
$\kom$
are $3\times 3$ real traceless matrices.
The matrix $\kop$ is antisymmetric,
and the other three are symmetric.
The remaining trace component $\ktr$
represents a single real coefficient
and is associated with isotropic violations.

The coefficients in
$\kem$, $\kop$, and $\ktr$
mimic a small distortion in the
spacetime metric,
resulting in a distorted version of the
usual electrodynamics.
In contrast,
the coefficients $\kep$ and $\kom$
break the usual two-fold degeneracy
that occurs in electrodynamics,
causing light to propagate as the
superposition of two modes that
differ in speed and polarization.
This causes birefringence
and results in a change in the net
polarization of light as it propagates.
Searches for birefringence in light 
from astrophysical sources
have resulted in stringent constraints
at the level of $10^{-32}$ or less
on the 10 coefficients in $\kep$ and $\kom$
\cite{km-bire}.
Consequently,
resonator experiments normally focus
on the 8 coefficients in
$\kop$ and  $\kem$,
which do not cause birefringence.
The isotropic coefficient $\ktr$
is not usually considered because
it is doubly suppressed.
However, in principle,
resonator experiments
can test all 19 coefficients.

\subsection{Resonator experiments}
\label{resonators}

Equation \rf{const} suggests that the
effects of Lorentz violation are
similar to those of linear media.
This analogy provides an intuitive
understanding of the basic principle
behind resonant-cavity experiments.
The matter effects from the $\ep$
matrices generally depend on the
orientation of the media
within the cavity.
However, 
since the location and orientation of
media are typically fixed with 
respect to the apparatus, 
the frequency does not change
with changes in the orientation or
velocity of the resonator.
In contrast,
the $\ka$ matrices can be viewed
as constant background fields 
pervading all of space.
The cavities are immersed in these 
background fields,
and changing the orientation
or velocity of the cavity with respect
to these fields can lead to
a change in resonant frequency.

To test for these effects,
experiments search for small
variations in resonant frequencies
with changes in orientation or velocity.
Rotations of the resonator
are normally achieved through either 
the sidereal motion of the Earth
or more actively through
the use of turntables.
Experiments monitor the frequency,
searching for rotation-violating
Michelson-Morley-type signals.
To date, this method has
yielded sensitivity to
parity-even coefficients only.
At present, $\kem$ is constrained
at the level of $\sim 10^{-16}$ by
Michelson-Morley techniques \cite{cavities}.

Sensitivity to the parity-odd $\kop$
has only been obtained through
Kennedy-Thorndike tests,
resulting in less stringent constraints.
The reason for this stems from the
fact that frequency is a
parity-even quantity.
In parity-symmetric resonators,
parity-odd violations can only affect
the frequency if they contribute 
in conjunction with another
parity-odd quantity.
Boost effects allow for this
since they involve a parity-odd velocity.
As a result,
Kennedy-Thorndike effects are usually
suppressed by a factor of
$\be\sim 10^{-4}$, 
the typical velocity of the apparatus.
Consequently,
current constraints on parity-odd
$\kop$ coefficients are near $10^{-12}$.

Similar symmetry arguments 
apply to the isotropic violations
associated with $\ktr$.
Isotropic effects are
difficult to observable,
but $\ktr$ does cause observable
boost violations.
However,
arguments similar to those given above
imply that these effects
enter suppressed by two factors of velocity,
giving a suppression factor
of $\sim 10^{-8}$ 
in parity-symmetric experiments.
While searching for these effects
in resonators is feasible,
current bounds on this coefficient
use other techniques \cite{sher}.

For resonators,
the effects of Lorentz violation
are characterized by the
leading-order shifts in 
resonant frequencies,
given by the generic expression
\begin{align}
  \fr{\de\nu}{\nu_0} =\ & 
  (\Mde)^{jk}(\kde)^{jk}
  +(\Mhb)^{jk}(\khb)^{jk}
  \notag \\  & 
  +(\Mdb)^{jk}(\kdb)^{jk} \ ,
  \label{dnu}
\end{align}
where
$(\Mde)^{jk}$,
$(\Mhb)^{jk}$, and
$(\Mdb)^{jk}$
are experiment-dependent factors.
Typically one begins an analysis by
calculating these dimensionless 
factors in a frame that is fixed 
to the resonator.
In this frame,
the $\cal M$ matrices are 
experiment-specific numerical constants.
In contrast,
the $\ka$ matrices are constant
only in inertial frames.
By convention,
a standard Sun-centered
inertial frame is used,
and all measurements are
reported in terms of coefficients 
in this frame.
A coordinate transformation is 
used to relate the 
resonator-frame $\ka$ matrices
to constant Sun-frame matrices.
This transformation introduces
the orientation and velocity
dependence that constitute
the signals for Lorentz violation.
Neglecting boost effects,
the resonator-frame and Sun-frame 
$\ka$'s are related by a rotation.
This implies that the 
unsuppressed Michelson-Morley-type
sensitivity to a particular 
Sun-frame $\ka$ matrix
is completely determined by the
corresponding resonator-frame 
$\cal M$ matrix.
For example,
an experiment with nonzero $\Mdb$
would be sensitive to rotational
effects associated with a nonzero $\kdb$.
In contrast,
zero $\Mdb$ implies that only
suppressed boost effects arise
from nonzero $\kdb$.

The
$\cal M$ matrices can be 
calculated perturbatively
in terms of the fields in the
absence of Lorentz violation,
$\mbf{E}_0$, $\mbf{D}_0$, 
$\mbf{B}_0$, and $\mbf{H}_0$
\cite{km}:
\begin{align}
  (\Mde)^{jk} &= -\frac{1}{4U}
  \int d^3x\ \Re\ (E_0^*)^j(E_0)^k \ ,
  \notag \\
  (\Mhb)^{jk} &= \phantom{-}\frac{1}{4U}
  \int d^3x\ \Re\ (B_0^*)^j(B_0)^k \ ,
  \label{Ms} \\
  (\Mdb)^{jk} &= -\frac{1}{2U}
  \int d^3x\ \Re\ (E_0^*)^j (B_0)^k \ ,
  \notag
\end{align}
where 
$U=\frac14\int d^3x\ 
( \mbf{E}_0^*\cdot\mbf{D}_0
+ \mbf{B}_0^*\cdot\mbf{H}_0 )$
is the time-averaged energy
stored in the resonator.
In what follows,
it will be useful to have a
birefringent decomposition 
of these matrices,
\begin{align}
  \Mep&=\Mde+\Mhb -\Frac13{\rm Tr}(\Mde+\Mhb)\ , \notag \\
  \Mem&=\Mde-\Mhb -\Frac13{\rm Tr}(\Mde-\Mhb)\ , \notag \\
  \Mop&=\Frac12(\Mdb-\Mdb^T)\ ,  \label{Mts}\\
  \Mom&=\Frac12(\Mdb+\Mdb^T) -\Frac13{\rm Tr}\Mdb\ , \notag \\
  \Mtr\ &={\rm Tr}(\Mde-\Mhb)=-1\ . \notag
\end{align}
These $\cal\widetilde M$ matrices
characterize the dependence on 
the $\tilde\ka$ matrices through an
expression analogous to Eq.\ \rf{dnu}.
Here we want to explore sensitivity
to parity-odd violations,
so our primary focus will be
on $\Mop$ and $\Mom$ matrices.

Note that the $\cal\widetilde M$
matrices are calculated using
conventional solutions.
Therefore,
to determine the effects of
Lorentz violation on 
resonator frequencies,
we need only to solve for the
fields in the Lorentz-invariant case.
Consequently,
we drop the subscript 0 on
all fields in what follows,
with the understanding that we
are working within the usual
Lorentz-invariant electrodynamics,
and all fields are conventional.

\section{parity-breaking resonators}
\label{parity-breaking}

Mathematically,
the reason parity-odd Lorentz violations
do not typically contribute at
unsuppressed levels is because 
the solutions can be 
split into solutions of 
definite parity.
In parity-symmetric cavities
with parity-conserving media,
the boundary conditions and
the Maxwell equations
normally admit conventional 
nondegenerate resonances
of the form,
$\mbf{E}_{\pm}(\mbf{x})
=\pm\mbf{E}_{\pm}(-\mbf{x})$,
$\mbf{B}_{\pm}(\mbf{x})
=\mp\mbf{B}_{\pm}(-\mbf{x})$.
The result is that $\Mdb$ in
Eq.\ \rf{Ms} vanishes,
since
$(E_{\pm}^*)^j(\mbf{x}) (B_{\pm})^k(\mbf{x})
=-(E_{\pm}^*)^j(-\mbf{x}) (B_{\pm})^k(-\mbf{x})$.
The result is a zero $\Mdb$,
which implies no sensitivity
to parity-odd Lorentz violations.
So,
in order to access parity-odd violations,
we should construct resonators
that admit solutions of indefinite parity.

Resonators could be constructed
that break parity symmetry
by using asymmetric geometries
or by introducing
parity-breaking media.
Below we demonstrate this
with an explicit example,
but we first show that,
in either case,
one cannot achieve unsuppressed
sensitivity to certain combinations
of Lorentz violations using 
a single lossless resonator.

We begin by noting
that the average flow of 
electromagnetic energy
within a volume $V$ can be split
into terms representing 
the energy flowing through
the surface of $V$ 
and the energy from 
sources and sinks within $V$.
The explicit expression,
in terms of the harmonic Poynting vector
$\mbf{S}=\half \Re\ \mbf{E}^*\times\mbf{H}$,
is given by the integral identity
\beq
\int_V \mbf{S}\ d^3x 
=\oint_{\prt V} \mbf{x}\ 
\mbf{S}\cdot d\mbf{a}
-\int_V \mbf{x}\ 
\mbf\nabla\!\cdot\mbf{S}\ d^3x \ .
\label{ident}
\eeq
For harmonic fields,
the source term vanishes since
$\mbf\nabla\!\cdot\mbf{S}=0$
in regions without current \cite{jackson}.
This is simply the statement that
there are no sources or sinks
of energy within the resonator.
This leaves the surface term
and the energy flowing through $\prt V$.
This term vanishes immediately
if the fields are sufficiently
confined to the interior of $V$. 
It also vanishes if we impose
perfect-conductor boundary conditions
at $\prt V$.
In this case, 
$\mbf{E}$ is perpendicular to 
the surface $\prt V$,
so $\mbf{S}$ is parallel to $\prt V$.
This implies that, on average,
no energy is exchanged with any
portion of the conductor,
and the average-energy-flow
lines are confined to
the volume of the cavity.
Equation \rf{ident} then implies that 
$\int_V \mbf{S}\ d^3 x=0$.
It follows that
\beq
\int_V \Re\ (\mbf{E}^*\otimes\mbf{H}
- \mbf{H}\otimes\mbf{E}^*)\ d^3x = 0\ .
\label{Sident}
\eeq
This $3\times 3$ antisymmetric
matrix equation places three
real constraints on the
$\cal M$ matrices.
This implies that,
for a given lossless resonator,
regardless of geometry,
there are at least three
combinations of coefficients 
for Lorentz violation that are
inaccessible at unsuppressed levels.

As an example,
consider a cavity filled with a 
frequency-independent medium.
In this case, 
the $\ep$ constituent matrices are real,
and the above discussion implies
\begin{align}
  2\, (\Mde)\cdot(\edb)
  -2\, (\edb)^T\!\cdot(\Mde)
  & \notag \\
  -(\Mdb)\cdot(\ehb)
  +(\ehb)\cdot(\Mdb)^T &= 0 \ ,
  \label{constraint}
\end{align}
assuming the the constituent matrices
$\ehb$ and $\edb$ are uniform 
throughout the volume $V$.
This matrix equation places 
three constraints on the 
$\cal M$ matrices,
implying that three combinations of
$\ka$ matrices are inaccessible.
A particularly relevant simple case
is a cavity containing a 
simple magnetic medium
with $\edb=0$  and $\ehb=\mu^{-1}$,
where $\mu$ is a homogeneous
isotropic permeability.
Equation \rf{constraint}
then implies that
the antisymmetric component 
of $\Mdb$ vanishes.
Consequently, $\Mop$ is zero
and $\kop$ no longer contributes
to the fractional frequency shift
at unsuppressed levels.
The conclusion is that
resonators incorporating
simple isotropic magnetic media
have no sensitivity to
nonbirefringent parity-odd violations.
Sensitivity to the three
components of $\kop$
can only be obtained by
the introduction
of more complicated
magnetic materials.
This is of particular interest
because the 3 coefficients in
$\kop$ are the least constrained
of the 18 anisotropic coefficients
for Lorentz violation.

\section{numerical example}
\label{example}

In this section,
we give a numerical example
of a resonant cavity
with parity-breaking geometry.
A numerical method for solving
the Maxwell equations in
curvilinear coordinates is given,
and used to illustrate some of
the conclusions of the previous section.

\subsection{Geometry}
\label{geometry}

\begin{figure}
  \includegraphics[width=0.3\textwidth]{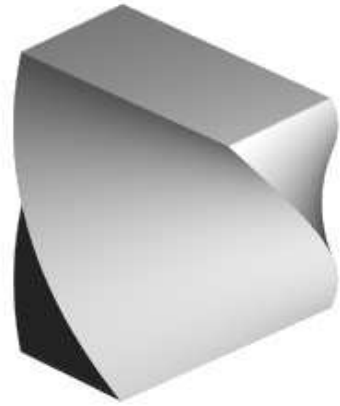}
  \vskip-5pt
  \caption{\label{cav}
    Example of a parity-breaking geometry.
    The object shown represents the
    cavity volume.}
\end{figure}

One way to ensure a
breakdown of parity symmetry
is to introduce a net
chirality in the cavity geometry.
We do this here by considering
a helical cavity,
as illustrated in Fig.\ \ref{cav}.
While we will assume that 
the cavity is empty,
the technique described here is readily 
adapted to cases involving linear media.

The geometry of this cavity 
can be characterized using
helical coordinates 
$x^a$, $a=1,2,3$,
related to standard cartesian coordinates
$x^j$, $j=x,y,z$, through
$x^1=x^x\cos\al x^z - x^y\sin\al x^z$,
$x^2=x^x\sin\al x^z + x^y\cos\al x^z$,
$x^3=x^z$.
Here we consider a cavities
with perfectly conducting boundaries at 
$x^a= \pm X^a$,
where $X^a$ are positive 
constants that specify the
cross-sectional and length
dimensions of the cavity.
The parameter $\al$ determines
the amount of rotation in the cavity
about the $x^3$ central axis.
For example, 
in Sec.\ \ref{results} we take
$X^1=1/2$, $X^2=1$, $X^3=1$,
and $\al=45^\circ$.
This gives a cavity with
a rectangular cross section 
and a quarter left-handed
turn from end to end,
as in Fig.\ \ref{cav}.

In curvilinear coordinates,
the conventional Maxwell equations
take the form
\begin{alignat}{2}
  g^{-1/2} \ \ep^{abc}\prt_b E_c + \prt_0 B^a &= 0 \ , &
  \quad \nabla_a B^a = 0 \ , \\
  g^{-1/2} \ \ep^{abc}\prt_b B_c - \prt_0 E^a &= 0 \ , &
  \quad \nabla_a E^a = 0 \ ,
\end{alignat}
where $E^a$ and $B^a$ are contravariant
field components, and
$E_a=g_{ab}E^b$ and $B_a=g_{ab}B^b$
are covariant components.
Here, $g_{ab}$ is the metric
in curvilinear coordinates,
and $\nabla_a$ is the associated
covariant derivative.
Note that the determinant $g=\det g_{ab} = 1$
in the helical coordinates used here.

One advantage to using 
curvilinear coordinates
is that the boundary conditions
become relatively simple.
Perfect-conductor boundary conditions imply
that $\mbf E$ is perpendicular and 
$\mbf B$ parallel to the conducting
surfaces of the cavity.
As an example,
consider a conducting boundary
whose surface is represented 
by constant $x^1$.
The contravariant basis vector
$\mbf e^1$ is perpendicular to this surface,
and covariant vectors
$\mbf e_2$ and $\mbf e_3$
are parallel to the surface.
So,
we must have
$\mbf E = E_1 \mbf e^1$ and
$\mbf B = B^2 \mbf e_2 + B^3 \mbf e_3$
at this boundary.
In our case, this implies
$E_{1,2}$ vanish at the 
ends ($x^3=\pm X^3$),
$E_{2,3}=0$ at $x^1=\pm X^1$,
and 
$E_{3,1}=0$ at $x^2=\pm X^2$.
For $\mbf B$,
we get vanishing
$B^3$ on the ends,
$B^1=0$ at $x^1=\pm X^1$,
and 
$B^2=0$ at $x^2=\pm X^2$.

\subsection{Discrete solutions} \label{discrete}

In order to show that the 
chiral geometry described above
does in fact produce sensitivity
to parity-odd Lorentz violations,
we perform a numerical analysis
of the its lowest-frequency resonances.
Finite-difference-time-domain (FDTD)
methods \cite{fdtd}
provide a straightforward
procedure for estimating
the $\cal M$ matrices over a
range of frequencies.
In this section,
we develop a FDTD procedure
for curvilinear coordinates.

We begin by defining discrete time by
taking $t_N= \de t \cdot N$,
where $\de t$ is a small time
interval, and $N$ is an integer.
The discrete fields are then taken as
$\mbf E_N=\mbf E(t_N)$ and 
$\mbf B_N=\mbf B(t_N-\de t/2)$.
This leads to discrete Maxwell equations:
\begin{align}
  (B^a)_{N+1}&\simeq(B^a)_N
  -\de t\ \ep^{abc}(\prt_b E_c)_N \ , 
  \label{Bn}\\
  (E^a)_{N+1}&\simeq(E^a)_N
  +\de t\ \ep^{abc}(\prt_b B_c)_{N+1} \ ,
  \label{En}
\end{align}
where we have assumed $g=1$.
This result allows us
to ``leapfrog'' through time
by iteratively
applying Eq.\ \rf{Bn}
followed by \rf{En}.

For spatial dimensions,
we construct a grid in 
helical coordinates,
\begin{alignat}{2}
  (x^1)_J &=&\ J\cdot\de x^1 &- X^1\ ,\notag \\
  (x^2)_K &=&\ K\cdot\de x^2 &- X^2\ ,\\
  (x^3)_L &=&\ L\cdot\de x^3 &- X^3\ ,\notag
\end{alignat}
where $\de x^a$ are small 
spatial intervals, 
$J,K,L$ are integers,
and $-X^a$ represent the
low edges of the cavity.
We then construct a pair
of lattices containing
field values $(E_a)_{NJKL}$ and 
$(B^a)_{NJKL}$ defined
at these spatial points.

In order to apply
Eqs.\ \rf{Bn} and \rf{En},
we need estimates for the
spatial derivatives.
Whenever possible,
we use the symmetric forms
\begin{align}
  (\prt_1 f)_{NJKL} &\simeq 
  [f_{N(J+1)KL}-f_{N(J-1)KL}]/(2\de x^1)\ , 
  \notag\\
  (\prt_2 f)_{NJKL} &\simeq 
  [f_{NJ(K+1)L}-f_{NJ(K-1)L}]/(2\de x^2)\ ,
  \label{deriv} \\
  (\prt_3 f)_{NJKL} &\simeq 
  [f_{NJK(L+1)}-f_{NJK(L-1)}]/(2\de x^3)\ .
  \notag
\end{align}
These can be used at 
each of the interior nodes,
but boundary nodes
must be treated more carefully since
derivatives \rf{deriv} are not
always defined at these points.
Also, some care must be taken to
ensure that the boundary conditions 
are satisfied at these nodes.

The method proceeds by
stepping the $\mbf B$ field
forward in time 
using Eqs.\ \rf{Bn} and
\rf{deriv} for interior nodes.
Next we propagate the boundary nodes.
Here we illustrate the procedure
for boundary nodes on the 
$J=0,\ x^1=-X^1$ surface.
The generalization to other boundary
surfaces is straightforward.

For $J=0$ boundary nodes,
the boundary conditions
imply vanishing
$E_2$, $E_3$, and $B^1$.
To propagate $\mbf B$ at one
of these nodes using Eq.\ \rf{Bn},
we need the partial derivatives 
$\prt_a E_b$ for $a\neq b$.
Since $E_{2,3}=0$ on this surface,
the derivatives
$\prt_2 E_3$ and $\prt_3 E_2$ vanish.
This implies that $B^1$ remains zero
provided that it vanished to begin with,
as required by the boundary conditions.
The derivatives
$\prt_2 E_1$ and $\prt_3 E_1$
can be estimated using the
symmetric form \rf{deriv}.
In contrast,
Eq.\ \rf{deriv} fails for
$\prt_1 E_2$ and $\prt_1 E_3$,
since it would require field
values at nodes outside
of the cavity.
So, in these cases
we use the one-sided derivative,
\beq
(\prt_1 E_{2,3})_{NJKL} \big|_{J=0} 
\simeq (E_{2,3})_{N(J+1)KL}
/ \de x^1 \big|_{J=0} \ ,
\eeq
where we take advantage of
the boundary conditions
$E_{2,3}=0$ at $J=0$.
We now have estimates for
all six spatial derivatives
$\prt_a E_b,\ a\neq b$ at
these nodes and
can use Eq.\ \rf{Bn} to
propagate $\mbf B$ at
this boundary.
The other boundary surfaces
are then propagated one
step in time using
similar methods.

Next, we propagate
$\mbf E$ at interior points
using Eqs.\ \rf{En} and \rf{deriv}.
Again, we will illustrate the procedure
for boundary nodes by considering
the $x^1=-X^1,\ J=0$ surface.
Since $E_2$ and $E_3$ vanish
on this surface,
we only need to calculate the
change in $E_1$.
However, since Eq.\ \rf{En}
propagates contravariant components,
some care is needed in developing
a procedure that updates $E_1$,
but leaves $E_2$ and $E_3$ unaltered.
We do this by noting that
\begin{align}
  &(E^1)_{(N+1)}-(E^1)_N
  \notag\\ &\quad\quad\quad 
  = g^{11}\left((E_1)_{(N+1)}-(E_1)_N\right)
  \\ &\quad\quad\quad 
  \simeq \de t\ \left( (\prt_2B_3)_{N+1}
    - (\prt_3B_2)_{N+1}\right) \ ,
  \notag
\end{align}
where we have used the fact
that $E_{2,3}=0$ to write
$E^1=g^{1a}E_a=g^{11}E_1$.
Using this result,
we can step $E_1$ fields in time
at $J=0$ nodes with the relation
\begin{align}
  &(E_1)_{(N+1)} - (E_1)_N
  \notag\\ &\quad\quad\quad
  \simeq \de t\ \left( (\prt_2B_3)_{N+1}
    - (\prt_3B_2)_{N+1}\right)/g^{11} \ .
\end{align}
Here, 
Eq.\ \rf{deriv} is used
to estimate spatial derivatives
without difficulty.
Again,
the other boundary surfaces
are propagated using the
generalization of this method.

By repeating the above procedure,
we can propagate the $\mbf B$ 
and $\mbf E$ fields 
in time indefinitely.
Note that at each time step,
the $\mbf E$ ($\mbf B$) fields 
at a given node depend on the
prior $\mbf E$ ($\mbf B$) fields 
at that node and the prior 
$\mbf B$ ($\mbf E$) fields
at adjacent nodes.
This implies that
$\mbf E$ and $\mbf B$ need
not be defined at every node,
and we may adopt a
lattice of fields in which
$\mbf E$ and $\mbf B$ are
only defined at alternate nodes.
For example,
in this work we take
$\mbf E$ defined at nodes
with $J+K+L=\mbox{even}$,
and $\mbf B$ defined at nodes
with $J+K+L=\mbox{odd}$,
forming two interlaced
$\mbf E_{NJKL}$ and
$\mbf B_{NJKL}$ lattices.
There is nothing preventing us
from defining
$\mbf E$ and $\mbf B$
at each node,
but,
in doing so,
the calculation would
essentially decouple into
the propagation of two
independent sets of fields
like the ones used here,
doubling the amount of
information that is necessary.

A similar observation in the cartesian
case led to a ``Yee cell''
in which different field components
are defined at different spatial points
\cite{fdtd}.
In our case,
the mixing of components resulting
from the raising and lowering
of spatial indices by way of
the metric makes the usual
Yee method impractical.

To initialize the calculation,
we must first seed the cavity
with divergenceless fields
satisfying the boundary conditions.
A convenient set of initial fields
is obtained by taking
the expressions for the usual 
transverse-magnetic (TM)
and transverse-electric (TE)
$\mbf B$ fields associated
with a rectangular cavity
with $\al=0$ and
making the substitutions
$\{B^x,B^y,B^z\}\rightarrow\{B^1,B^2,B^3\}$
and
$\{x^x,x^y,x^z\}\rightarrow\{x^1,x^2,x^3\}$.
For simplicity,
we simply set the initial 
$\mbf E$ fields to zero.
The resulting initial fields
obey the correct boundary conditions
and can be shown to be divergenceless.
Once the fields are set to these
valid initial values,
we can then propagate the fields
in time using the above procedure.

\subsection{Results} 
\label{results}

\begin{figure}
  \vskip3pt
  \centerline{\includegraphics[width=0.45\textwidth]{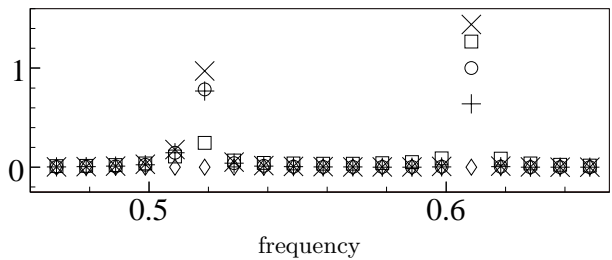}}
  \centerline{frequency}
  \caption{\label{plot}
    Numerically-determined
    lowest-frequency resonances
    for a cavity with 
    $X^1=1/2$, $X^2=1$, $X^3=1$,
    and $\al=45^\circ$.
    Circles represent
    the energy $U$ versus 
    frequency.
    Magnitudes of parity-odd 
    $\cal\widetilde M$
    matrices are plotted as 
    $100 \times U \times |\Mop|$ (diamonds)
    and 
    $100 \times U \times|\Mom|$ (squares).
    For comparison,
    we also show 
    $5\times U\times |\Mep|$ ($\times$ signs)
    and
    $5\times U\times |\Mem|$ ($+$ signs).
    All values are normalized
    so that the maximum energy $U_{max}=1$
    in arbitrary units.
  }
\end{figure}

We next apply the method
described in the previous section
to the cavity shown in Fig.\ \ref{cav}.
The dimensions of the cavity,
in arbitrary spacetime units,
are taken as
$X^1=1/2$, $X^2=1$, and $X^3=1$.
Taking $\al=45^\circ$ gives a quarter
left-handed twist as shown in the figure.
Applying the initialization method
described above,
we set initial-field values using
the conventional expressions for the
magnetic fields associated with the
TM$_{110}$ mode for the analogous
rectangular cavity with $\al=0$.

We use a spatial lattice 50 nodes
wide in each of the three helical
coordinates.
We take a total time
interval of 100 and
calculate a total
of 20,000 time steps.
In order to reduce the amount of
data saved to disk,
we only record the field values for
every fiftieth step.
A fast fourier transform is
performed on the saved field values,
at each spatial node,
yielding frequency-domain data.
Using these,
we determine the energy $U$ and
the $\cal\widetilde M$ matrices,
in cartesian coordinates,
as a function of frequency.
The results near the two lowest
resonances are shown in Fig.\ \ref{plot}.
The matrix magnitudes,
$|{\cal M}|=
\sqrt{{\cal M}^{jk} {\cal M}^{jk}}$,
for $\Mep$, $\Mem$, $\Mop$, and $\Mom$
are shown in the figure.

As expected,
this parity-breaking configuration
gives rise to a nonzero $\Mom$
at both of the resonances shown
in Fig.\ \ref{plot},
demonstrating sensitivity to the
parity-odd violations associated with $\kom$.
Furthermore,
we find that $\Mop=0$ to within the
errors of the calculation.
This confirms the predictions of
Sec.\ \ref{parity-breaking},
showing that
while sensitivities
to parity-odd Lorentz violations
are possible,  sensitivity
to the nonbirefringent parity-odd
violations cannot be achieved
in resonators with simple 
isotropic magnetic media.
We also note that $|\Mom|$
appears to be significantly
smaller in the lower-frequency resonance,
suggesting that sensitivities to
parity-odd Lorentz violations are
likely to be strongly dependent on 
the resonant mode excited in the cavity.

For both of the resonances in
Fig.\ \ref{plot},
we find nonzero $\Mep$ and $\Mem$ matrices,
demonstrating sensitivity to
the violations associated with
$\kep$ and $\kem$,
as in parity-even cavities.
We note that the sensitivities
to parity-odd violations
in this example 
are larger by roughly a 
order of magnitude relative to
parity-even violations.
This geometric suppression
shows that even in resonators
with significant parity asymmetries,
sensitivity to parity-odd
violations may be small
compared to those for
parity-even violations.
Nevertheless,
this demonstrates the
potential for at least 
a thousand-fold improvement
in sensitivity to parity-odd
Lorentz violations,
assuming cavities of this type
could be constructed and
achieve stabilities comparable
to their symmetric counterparts.

\section{summary and outlook}
\label{discussion}

At present,
resonant-cavity experiments
have achieved sensitivities
near $10^{-16}$ to 
parity-even coefficients
for Lorentz violation
\cite{cavities}.
The parity-odd coefficients
enter through suppressed boost effects, 
resulting in constraints that
are larger by approximately
four orders of magnitude.
Here we have shown that resonant cavities
that do not respect parity symmetry
can provide unsuppressed sensitivity 
to parity-odd Lorentz violations.
Parity asymmetries can be introduced
through the geometry of the cavity
or by incorporating parity-breaking media.

In principle, the parity-odd
coefficients $\kop$ and $\kom$
can cause observable violations
of rotation symmetry in 
parity-breaking cavities,
leading to improved sensitivities.
In particular,
this idea could be used
to make significantly
tighter constraints on
the nonbirefringent
$\kop$ coefficients.
However,
some thought must go into the
design of a resonator to 
ensure sensitivity to $\kop$.
In Sec.\ \ref{parity-breaking},
we have shown that a
given resonator
will be insensitive to certain 
combinations of coefficients.
More specifically,
we have shown that sensitivity
to the three coefficients in $\kop$
is not possible in 
cavities incorporating
only simple isotropic
magnetic media.

Better sensitivities to $\kop$
could be achieved in resonators
utilizing a combination of
anisotropic magnetic media,
with nondegenerate $\ehb$,
in conjunction with
asymmetric geometries
or by using  parity-violating media 
with nonzero $\edb$.
Chiral media \cite{chiral}
provide another interesting possibility.
Assuming stabilities comparable
to those in current experiments,
parity-asymmetric resonators
have the potential to improve 
the constraints on $\kop$ coefficients 
by four orders of magnitude
by circumventing the boost suppression
associated with Kennedy-Thorndike tests.

Resonators of this type could also be used
to place improved laboratory-bounds
on the five parity-odd coefficients in $\kom$.
While cavity tests are not likely to
achieve the same kind of 
sensitivities that are
obtained in searches for birefringence,
these experiments could provide a valuable
laboratory-based check on
astrophysical bounds.
As illustrated in Sec.\ \ref{example},
sensitivities to $\kom$ can be improved
simply by using parity-breaking geometries.

Finding geometries and media
that maximize sensitivities
to parity-odd effects remains
an interesting open problem.
The construction of high-Q
asymmetric cavities may also pose
a technological challenge.
However,
development of parity-breaking resonators
would provide another avenue
for high-precision tests of
Lorentz invariance that would compliment
the current parity-symmetric experiments.
They have the potential to yield
significant improvements in
sensitivities to parity-odd
Lorentz violation
and could rival the best tests
in any sector.


\begin{thebibliography}{99}

\bibitem{mm}
  A.A.\ Michelson and E.W.\ Morley,
  Am.\ J.\ Sci.\ {\bf 34}, 333 (1887);
  Phil.\ Mag.\ {\bf 24}, 449 (1887).

\bibitem{kt}
  R.J.\ Kennedy and E.M.\ Thorndike,
  Phys.\ Rev.\ {\bf 42}, 400 (1932).

\bibitem{cavities}
  J.\ Lipa {\it et al.}, Phys.\ Rev.\ Lett.\ \bf 90\rm, 060403 (2003);
  H.\ M\"uller {\it et al.}, Phys.\ Rev.\ Lett.\ \bf 91\rm, 020401 (2003);
  H.\ M\"uller {\it et al.}, Phys.\ Rev.\ D \bf 67\rm, 056006 (2003); 
  Phys.\ Rev.\ D \bf 68\rm, 116006 (2003);
  P.\ Wolf {\it et al.}, Gen.\ Rel.\ Grav.\ \bf 36\rm, 2352 (2004);
  P.\ Wolf {\it et al.}, Phys.\ Rev.\ D \bf 70\rm, 051902 (2004);
  H.\ M\"uller, Phys.\ Rev.\ D \bf 71\rm, 045004 (2005);
  P.L.\ Stanwix {\it et al.}, 
  Phys.\ Rev.\ Lett.\ \bf 95\rm, 040404 (2005);
  Phys.\ Rev.\ D {\bf 74},  081101 (2006);
  S.\ Herrmann {\it et al.}, Phys.\ Rev.\ Lett.\ \bf 95\rm, 150401 (2005);
  P.\ Antonini {\it et al.}, Phys.\ Rev.\ A \bf 71\rm, 050101 (2005).
  
\bibitem{ks}
  V.A.\ Kosteleck\'y and S.\ Samuel,
  Phys.\ Rev.\ D {\bf 39}, 683 (1989);
  {\bf 40}, 1886 (1989);
  Phys.\ Rev.\ Lett.\ {\bf 63}, 224 (1989);
  {\bf 66}, 1811 (1991).

\bibitem{kp}
  V.A.\ Kosteleck\'y and R.\ Potting,
  Nucl.\ Phys.\ B {\bf 359}, 545 (1991);
  Phys.\ Lett.\ B {\bf 381}, 89 (1996);
  Phys.\ Rev.\ D {\bf 63}, 046007 (2001);
  V.A.\ Kosteleck\'y, M.\ Perry, and R.\ Potting,
  Phys.\ Rev.\ Lett.\ {\bf 84}, 4541 (2000).

\bibitem{ncqed}
  S.M.\ Carroll {\it et al.},
  Phys.\ Rev.\ Lett.\ {\bf 87}, 141601 (2001);
  Z.\ Guralnik, R.\ Jackiw, S.Y.\ Pi, and A.P.\ Polychronakos,
  Phys.\ Lett.\ B {\bf 517}, 450 (2001);
  C.E.\ Carlson, C.D.\ Carone, and R.F.\ Lebed,
  Phys.\ Lett.\ B {\bf 518}, 201 (2001);
  A.\ Anisimov, T.\ Banks, M.\ Dine, and M.\ Graesser,
  Phys.\ Rev.\ D {\bf 65}, 085032 (2002);
  I.\ Mocioiu, M.\ Pospelov, and R.\ Roiban,
  Phys.\ Rev.\ D {\bf 65}, 107702 (2002);
  M.\ Chaichian, M.M.\ Sheikh-Jabbari, and A.\ Tureanu,
  hep-th/0212259;
  J.L.\ Hewett, F.J.\ Petriello, and T.G.\ Rizzo,
  Phys.\ Rev.\ D {\bf 66}, 036001 (2002).
  
\bibitem{qg}
  R.\ Gambini and J.\ Pullin,
  in 
  V.A.\ Kosteleck\'y, ed.,
  \it CPT and Lorentz Symmetry, \rm
  World Scientific, Singapore, 1999;
  J.\ Alfaro, H.A.\ Morales-T\'ecotl, L.F.\ Urrutia,
  Phys.\ Rev.\ {\bf D66}, 124006 (2002);
  D.\ Sudarsky, L.\ Urrutia, and H.\ Vucetich,
  Phys.\ Rev.\ Lett.\ {\bf 89}, 231301 (2002);
  Phys.\ Rev.\ D {\bf 68}, 024010 (2003);
  G.\ Amelino-Camelia,
  Mod.\ Phys.\ Lett.\ A {\bf 17}, 899 (2002);
  Y.J.\ Ng,
  Mod.\ Phys.\ Lett.\ {\bf A18}, 1073 (2003);
  R.\ Myers and M.\ Pospelov,
  Phys.\ Rev.\ Lett.\ {\bf 90}, 211601 (2003);
  N.E.\ Mavromatos,
  hep-ph/0305215.
  
\bibitem{fn}
  C.D.\ Froggatt and H.B.\ Nielsen,
  hep-ph/0211106.

\bibitem{bj}
  J.D.\ Bjorken,
  Phys.\ Rev.\ D {\bf 67}, 043508 (2003).

\bibitem{cpt}
  For an overview see, for example,
  R.\ Bluhm, 
  Living Rev.\ Rel.\ {\bf 8}, 5 (2005);
  V.A.\ Kosteleck\'y, ed.,
  {\it CPT and Lorentz Symmetry},
  World Scientific, Singapore, 1999;
  {\it CPT and Lorentz Symmetry II},
  World Scientific, Singapore, 2002;
  {\it CPT and Lorentz Symmetry III},
  World Scientific, Singapore, 2005.

\bibitem{ck} 
  D.\ Colladay and V.A.\ Kosteleck\'y,
  Phys.\ Rev.\ D {\bf 55}, 6760 (1997);
  {\bf 58}, 116002 (1998).

\bibitem{kost}
  V.A.\ Kosteleck\'y,
  Phys.\ Rev.\ D {\bf 69}, 105009 (2004).

\bibitem{kb}
  Q.\ Bailey and V.A.\ Kosteleck\'y,
  Phys.\ Rev.\ D {\bf 70}, 076006 (2004).

\bibitem{tobar}
  M.E.\ Tobar {\it et al.}, Phys.\ Rev.\ D {\bf 71}, 025004 (2005).

\bibitem{km}
  V.A.\ Kosteleck\'y and M.\ Mewes,
  Phys.\ Rev.\ D {\bf 66}, 056005 (2002).

\bibitem{km-bire}
  V.A.\ Kosteleck\'y and M.\ Mewes,
  Phys.\ Rev.\ Lett.\ {\bf 87}, 251304 (2001);
  Phys.\ Rev.\ Lett.\ {\bf 97}, 140401 (2006).

\bibitem{cfj}
  S.M.\ Carroll, G.B.\ Field, and R.\ Jackiw, 
  Phys.\ Rev.\ D {\bf 41}, 1231 (1990).

\bibitem{photonth}
  M.P.\ Haugan and F.\ Kauffmann,
  Phys.\ Rev.\ D {\bf 52}, 3168 (1995);
  R.\ Jackiw and V.A.\ Kosteleck\'y,
  Phys.\ Rev.\ Lett.\ {\bf 82}, 3572 (1999);
  C.\ Adam and F.R.\ Klinkhamer,
  Nucl.\ Phys.\ B {\bf 657}, 214 (2003);
  H.\ M\"uller, C.\ Braxmaier, S.\ Herrmann, 
  A.\ Peters, and C.\ L\"ammerzahl,
  Phys. Rev. D {\bf 67}, 056006 (2003);
  T.\ Jacobson, S.\ Liberati, and D.\ Mattingly,
  Phys.\ Rev.\ D {\bf 67}, 124011 (2003);
  V.A.\ Kosteleck\'y, R.\ Lehnert, and  M.J.\ Perry, 
  Phys.\ Rev.\ D, {\bf 68}, 123511 (2003);
  V.A.\ Kosteleck\'y and A.G.M.\ Pickering,
  Phys.\ Rev.\ Lett.\ {\bf 91}, 031801 (2003);
  R.\ Lehnert, 
  Phys.\ Rev.\ D {\bf 68}, 085003 (2003);
  G.M.\ Shore, Nucl.\ Phys.\ B {\bf 717}, 86 (2005); 
  B.\ Altschul and V.A.\ Kosteleck\'y, Phys.\ Lett.\ B {\bf 628}, 106 (2005); 
  R.\ Bluhm and  V.A.\ Kosteleck\'y, Phys.\ Rev.\ D {\bf 71}, 065008 (2005);
  B.\ Altschul, hep-th/0609030.

\bibitem{sher}
  C.D.\ Carone, M.\ Sher, and M.\ Vanderhaeghen,
  Phys.\ Rev.D {\bf 74}, 077901 (2006).

\bibitem{ccexpt}
  L.R.\ Hunter {\it et al.},
  in 
  V.A.\ Kosteleck\'y, ed.,
  \it CPT and Lorentz Symmetry, \rm
  World Scientific, Singapore, 1999;
  V.A.\ Kosteleck\'y and C.D.\ Lane,
  Phys.\ Rev.\ D {\bf 60}, 116010 (1999);
  J.\ Math.\ Phys.\ {\bf 40}, 6245 (1999);
  D.\ Bear {\it et al.},
  Phys.\ Rev.\ Lett.\ {\bf 85}, 5038 (2000);
  D.F.\ Phillips {\it et al.},
  Phys.\ Rev.\ D {\bf 63}, 111101 (2001);
  M.A.\ Humphrey {\it et al.},
  physics/0103068;
  Phys.\ Rev.\ A {\bf 62}, 063405 (2000);
  F.\ Cane {\it et al.}, 
  Phys.\ Rev.\ Lett.\ {\bf 93}, 230801 (2004); 
  P.\ Wolf {\it et al.},
  Phys.\ Rev.\ Lett.\ {\bf 96}, 060801 (2006).

\bibitem{spaceexpt}
  R.\ Bluhm {\it et al.},
  Phys.\ Rev.\ Lett.\ {\bf 88}, 090801 (2002);
  Phys.\ Rev.\ D {\bf 68}, 125008 (2003).
  
\bibitem{hadrons}
  OPAL Collaboration, R.\ Ackerstaff {\it et al.}, 
  Z.\ Phys.\ C {\bf 76}, 401 (1997);
  BELLE Collaboration, K.\ Abe {\it et al.},
  Phys.\ Rev.\ Lett.\ {\bf 86}, 3228 (2001);  
  KTeV Collaboration, H.\ Nguyen, 
  in V.A.\ Kosteleck\'y, ed., \it CPT and Lorentz Symmetry II, \rm
  World Scientific, Singapore, 2002;
  FOCUS Collaboration, J.M.\ Link {\it et al.},
  Phys.\ Lett.\ B {\bf 556}, 7 (2003);  
  BaBar collaboration, B.\ Aubert {\it et al.}, 
  Phys.\ Rev.\ D {\bf 70}, 012007 (2004);
  Phys.\ Rev.\ Lett.\ {\bf 92}, 142002 (2004);
  hep-ex/0607103.

\bibitem{hadronth}
  D.\ Colladay and V.A.\ Kosteleck\'y, 
  Phys.\ Lett.\ B {\bf 344}, 259 (1995);
  Phys.\ Rev.\ D {\bf 52}, 6224 (1995); 
  Phys.\ Lett.\ B {\bf 511}, 209 (2001);
  V.A.\ Kosteleck\'y and R.\ Van Kooten, 
  Phys.\ Rev.\ D {\bf 54}, 5585 (1996);
  O.\ Bertolami {\it et al.},
  Phys.\ Lett.\ B {\bf 395}, 178 (1997);
  V.A.\ Kosteleck\'y,
  Phys.\ Rev.\ Lett.\ {\bf 80}, 1818 (1998);
  Phys.\ Rev.\ D {\bf 61}, 016002 (2000);
  {\bf 64}, 076001 (2001);
  N.\ Isgur {\it et al.},
  Phys.\ Lett.\ B {\bf 515}, 333 (2001);
  B.\ Altschul, hep-ph/0610324; hep-ph/0608094.

\bibitem{electrons}
  H.\ Dehmelt {\it et al.},
  Phys.\ Rev.\ Lett.\ {\bf 83}, 4694 (1999);
  R.\ Mittleman {\it et al.},
  Phys.\ Rev.\ Lett.\ {\bf 83}, 2116 (1999);
  G.\ Gabrielse {\it et al.},
  Phys.\ Rev.\ Lett.\ {\bf 82}, 3198 (1999);
  R.\ Bluhm {\it et al.},
  Phys.\ Rev.\ Lett.\ {\bf 82}, 2254 (1999);
  Phys.\ Rev.\ Lett.\ {\bf 79}, 1432 (1997);
  Phys.\ Rev.\ D {\bf 57}, 3932 (1998).  

\bibitem{electronth}
  R.\ Bluhm and V.A.\ Kosteleck\'y, 
  Phys.\ Rev.\ Lett.\ {\bf 84}, 1381 (2000);
  B.\ Heckel, in V.A.\ Kosteleck\'y, ed., 
  \it CPT and Lorentz Symmetry II, \rm
  World Scientific, Singapore, 2002;
  L.-S.\ Hou, W.-T.\ Ni, and Y.-C.M.\ Li, 
  Phys.\ Rev.\ Lett.\ {\bf 90}, 201101 (2003);
  B.\ Altschul, 
  Phys.\ Rev.\ D {\bf 72}, 085003 (2005);
  Phys.\ Rev.\ Lett.\ {\bf 96}, 201101 (2006);
  Phys.\ Rev.\ D {\bf 74}, 083003 (2006).

\bibitem{muons} 
  V.W.\ Hughes {\it et al.},
  Phys.\ Rev.\ Lett.\ {\bf 87}, 111804 (2001);
  R.\ Bluhm {\it et al.},
  Phys.\ Rev.\ Lett.\ {\bf 84}, 1098 (2000).

\bibitem{neutrinos}
  V.A.\ Kostelecky and M.\ Mewes, 
  Phys.\ Rev.\ D {\bf 69} 016005 (2004);
  {\bf 70} 031902 (2004); 
  {\bf 70} 076002 (2004);
  LSND Collaboration, L.B.\ Auerbach {\it et al.},
  Phys.\ Rev.\ D {\bf 72} 076004 (2005);
  M.D. Messier (Super Kamiokande), in V.A.\ Kosteleck\'y, ed., 
  \it CPT and Lorentz Symmetry III, \rm
  World Scientific, Singapore, 2005;
  B.J.\ Rebel and S.F.\ Mufson (MINOS), in V.A.\ Kosteleck\'y, ed., 
  \it CPT and Lorentz Symmetry III, \rm
  World Scientific, Singapore, 2005;
  T.\ Katori, A.\ Kostelecky, and R.\ Tayloe, 
  Phys.\ Rev.\ D {\bf 74} 105009 (2006).

\bibitem{higgs}
  D.L.\ Anderson, M.\ Sher, I.\ Turan,
  Phys.\ Rev.\ D {\bf 70}, 016001 (2004).

\bibitem{kb-gr}
  Q.\ Bailey and V.A.\ Kosteleck\'y,
  Phys.\ Rev.\ D {\bf 74}, 045001 (2006).

\bibitem{kp2}
  V.A.\ Kosteleck\'y and R.\ Potting,
  Gen.\ Rel.\ Grav.\ {\bf 37}, 1675 (2005).

\bibitem{kpo}
  V.A.\ Kosteleck\'y and R.\ Potting,
  Phys.\ Rev.\ D {\bf 51}, 3923 (1995).
    
\bibitem{kle}
  V.A.\ Kosteleck\'y and R.\ Lehnert,
  Phys.\ Rev.\ D {\bf 63}, 065008 (2001).
  
\bibitem{kong}
  J.A.\ Kong,
  \em Electromagnetic Wave Theory, \em
  Wiley, New York, 1990.

\bibitem{jackson}
  J.D.\ Jackson,
  \em Classical Electrodynamics, 3rd ed., \em 
  Wiley, New York, 1999. 

\bibitem{fdtd}
  K.\ Yee, IEEE Trans.\ Ant.\ Prop.
  {\bf 14}, 302 (1966).

\bibitem{chiral}
  See, for example, John Lekner,
  Pure Appl. Opt. {\bf 5}, 417 (1996).


  







\end{thebibliography}
\end{document}